\documentclass[12pt]{iopart}

\usepackage{graphicx}
\usepackage{hyperref}
\usepackage{amssymb}

\begin{document}

\title{Cavity cooling of an atomic array}

\author{O.S. Mishina}

\address{Theoretische Physik, Universit\"{a}t des Saarlandes, D-66041 Saarbr\"{u}cken, Germany}

\ead{omishina@physik.uni-saarland.de}


\begin{abstract}

While cavity cooling of a single trapped emitter was demonstrated, cooling of many particles in an array of harmonic traps needs investigation and poses a question of scalability. This work investigates the cooling of a one dimensional atomic array to the ground state of motion via the interaction with the single mode field of a high-finesse cavity. The key factor ensuring the cooling is found to be the mechanical inhomogeneity of the traps. Furthermore it is shown that the pumped cavity mode does not only mediate the cooling but also provides the necessary inhomogeneity if its periodicity differs from the one of the array. This configuration results in the ground state cooling of several tens of atoms within a few milliseconds, a timescale compatible with current experimental conditions. Moreover, the cooling rate scaling with the atom number reveals a drastic change of the dynamics with the size of the array: atoms are either cooled independently, or via collective modes. In the latter case the cavity mediated atom interaction destructively slows down the cooling as well as increases the mean occupation number, quadratically with the atom number. Finally, an order of magnitude speed up of the cooling is predicted as an outcome the optimization scheme based on the adjustment of the array versus the cavity mode periodicity.

\end{abstract}

\maketitle

\section{Introduction}

The possibility of trapping chains of atoms \cite{Gupta2007,Schleier-Smith2011,Brandt2010} and Wigner crystals  \cite{Herskind2009} in an optical cavity provides a new platform to study quantum optomechanics \cite{Stamper-Kurn2012,Ritsch2013}. The advantage of this system compared to other optomechanical platforms (i.e., micro- and nanometer scale mechanical oscillators) is the access to internal atomic degrees of freedom that can be used to tune the coupling and to manipulate the mechanical modes. The cavity does not only provide tailored photonic modes to interact with the atomic mechanical modes, it also alters the radiative properties of atoms giving rise to cavity mediated atom-atom interactions and collective effects. The combination of these ingredients results in a high degree of control of the optomechanical interface which has allowed, for example, the experimental observation of cavity nonlinear dynamics at a single photon level \cite{Gupta2007} and ponderomotive squeezing of light \cite{Brooks2012}. Such a platform, in which the optomechanical system includes multiple mechanical oscillators in the quantum regime globally coupled to the cavity field, shall eventually allow multipartite entanglement of distant atom motion \cite{Peng2002,Li2006}, hybrid light-motion entanglement \cite{Peng2002}, \cite{Cormick2013} and also engineering of spin-phonon coupling mediated by light when considering the atomic internal degrees of freedom.

An important problem on the way to reach the quantum optomechanical regime, is the cooling of the atomic mechanical modes to the ground state. Several techniques can be envisaged to prepare an atomic chain in the ground state of motion. One way is to prepare the atoms in the ground state of an optical lattice prior to coupling them to the cavity field. Sidebandanalysed resolved laser cooling can be used in this case \cite{Hamann1998}, but the implementation of Raman sideband cooling is restricted to atomic species with a suitable cycling transition. Another route, very powerful and experimentally convenient, is to use the cavity mode itself for cooling the atomic chain. It eliminates the need for additional preparation steps and allows reusing the same atoms multiple times. Moreover, it is not restricted to specific atomic species and can be potentially extended to the cooling of any polarizable object such as, for example, molecules \cite{Lev2008}. While the problem of cooling a single trapped particle in a cavity was explored theoretically \cite{Cirac1995a,Vuletic2001,Zippilli2005} and experimentally \cite{Leibrandt2009}, the simultaneous cooling of many particles forming an array poses the question of scalability. Cooling of an atomic array using a cavity mode was experimentally demonstrated in Ref. \cite{Schleier-Smith2011} where a single mode of the collective atomic motion was cooled close to the ground stare. The cooling rate of this unique collective mode was found to be proportional to the number of atoms in the array. A similar scaling was reported in a theoretical work for the case when a homogeneous cloud is first organized by the cavity potential and than collectively cooled \cite{Elsasser2003}.

A number of questions remain open on the protocol to cool down an array of atoms to the ground state inside a cavity. What is the role of the collective modes in the cooling dynamics of individual atoms? How do the cooling rates of individual atoms scale with the number of atoms in the array? What is the role of the lattice periodicity \emph{vs} the cavity mode period? What is the most efficient cooling scheme? This work provides the  answers to these questions. It shows that (i) cooling of a single collective mode is faster than the cooling of individual atoms, which is destructively suppresed due to collective effects, (ii) the cooling time for individual atoms increases non-linearly with the atom number, and (iii) the periodicity of the array plays a key role in the dynamics which can be used to optimize the cooling performance. Additionally it considers the limitations imposed by the spontaneous estimate outside of the cavity mode and shows the experimental feasibility of the cavity cooling of tens of atoms in the array. 

In order to address these questions, a theoretical model is developed describing the general configuration in which the cavity potential and the atomic array have different periodicity as, for example, implemented in ref. \cite{Schleier-Smith2011}. The key factor insuring the ground state cooling of all atoms via global coupling to the single cavity mode is found to be the mechanical inhomogeneity of the traps. The cavity mode itself is demonstrated to provide the nesessary inhomogeneity due to the effect of the cavity potential on the individual traps. This controlability makes the configuration of an atomic array coupled to the cavity with different periodicity an attractive platform for further investigation of a multimode quantum optomethanical interface. Additionally, the proposed cavity cooling scheme can be extended to the case of an array of micro- or nanometer scale mechanical oscillators, where strong optomechanical coupling was recently predicted \cite{Xuereb2012}. 

The paper is organized as follows. Section \ref{sec_model} summarizes the theoretical model and describes the physical mechanisms governing the cooling dynamics. In section \ref{sec_anal_res} we present the analytical results for the scaling of the cooling rates with the atom number. Section \ref{sec_num_res} compares numerical and analytical results for the cooling rates and the steady state mean phonon number per atom. The transition between two distinct regimes, when atoms interact independently or collectively with the cavity field, is reported. Also the destructive suppression of the cooling due to collective effects is demonstrated. In section \ref{sec_optimization} the role of the lattice periodicity \emph{vs} the cavity mode period is discussed and a possible way to speed up the cooling is suggested. Finally the effect of the spontaneous emission on the scaling of the steady state phonon number is analysed in section \ref{sec_spont_em} together with the experimental feasibility of the proposed cooling scheme. The conclusions are drawn in section \ref{sec_conclusion}.

\section{Summary of the model}
\label{sec_model}

\begin{figure}
\begin{center}
\includegraphics[width=0.5\columnwidth]{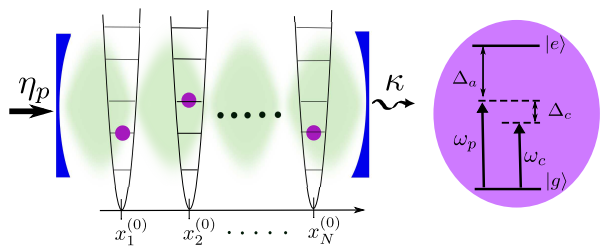}
\end{center}
\caption{The schematic representation of the system. $N$ individually trapped atoms are placed inside a cavity with resonant frequency $\omega_{c}$ and decay rate $\kappa$. Atoms have identical two-level structure and a resonance frequency $\omega_{eg}$. The cavity is pumped by an external laser, detuned by $\Delta_{c}=\omega_{p}-\omega_c$ from the cavity resonance frequency and by $\Delta_a=\omega_p-\omega_{eg}$ from the atomic transition frequency.}
\label{fig_schema}
\end{figure}

The system under investigation consists of two elements: (i) a one dimensional array of $N$ independently trapped atoms coupled to (ii) a quantum light field with wave number $k_c$ confined inside an optical cavity pumped by a monomode laser as presented in figure \ref{fig_schema}. The chain of two-level atoms is formed along the axis of the cavity where the atoms are confined in a deep optical lattice potential generated by an additional external classical field \cite{Gupta2007,Schleier-Smith2011}. The case of hopping and tunnelling of atoms between the different sites will be neglected. The trap array holding the neutral atoms may be experimentally implemented in various ways. In the works \cite{Gupta2007,Schleier-Smith2011} an extra cavity pump field resonant to the other cavity frequency was used to create a deep optical lattice. Alternatively, an optical lattice along the cavity can be created by two laser beams crossing each other at an angle inside the cavity or with the use of a spatial light modulator. Although the focus of this work is on the cooling of neutral atoms it is worth noticing that the generalization of the model for the case of of ions or other polarizable particles can be straightforwardly done.

The main mechanism behind the cavity cooling is the scattering process taking place when an atom absorbs a photon with pump frequency $\omega_p$ and then emits a photon back into the cavity with frequency $\omega_c$. If the pump frequency is lower than the cavity resonance frequency ($\Delta_c=\omega_p-\omega_c < 0$) and the difference is equal to the atomic trap frequency $\nu$, the atom will lose one vibration quantum, and the photon, eventually leaving the cavity, will carry this energy away. Such a cooling mechanism essentially relies on the interaction of atoms with the cavity field and assumes that the spontaneous emission into free space is negligibly small. This requires the cavity-to-free space scattering ratio to be much larger than one, which is reached when the single atom cooperativity (Purcell number) $c_r=\frac{g^2}{\kappa\gamma}$ is larger than one, regardless of the pump filed detuning from the atomic transition $\Delta_a=\omega_p-\omega_{eg}$ \cite{Vuletic2001}. It is achieved when a light-atom coupling strength $g$ is larger than the geometric average of the atomic natural linewidth $\gamma$ and the cavity decay rate $\kappa$.

We will focus on the regime in which the cavity field is far off-resonance from the atomic transition $|g\rangle \leftrightarrow |e\rangle$, such that the probability of an atomic excitation is negligibly small. Under the conditions  $|\Delta_a|\gg\gamma,\kappa,g\sqrt{N_{ph}}$, where $N_{ph}$ is the mean photon number in the cavity, the atomic internal degree of freedom can be adiabatically eliminated. In this case the coherent part of the optomechanical interaction between the cavity and atomic motion is described by the  effective Hamiltonian \cite{Domokos2003,Larson2008}:
\begin{eqnarray}
\label{Hammiltonian_spin_phot_phon}
H=&-&\hbar \left(\Delta_c-U_0 \sum_{i=1}^{N}\cos^2(k_c x^{(0)}_i+k_c\hat{x}_i) \right)\hat{A}^\dag \hat{A}
+\sum_{i=1}^{N}\left(\frac{m \nu^2\hat{x}^2_i}{2}+\frac{\hat{p}^2_i}{2m}\right)
\nonumber
\\
&+&i\hbar\left(\frac{•}{•}\eta_p\hat{A}^\dag-\eta_p^*\hat{A}\frac{•}{•}\right).
\end{eqnarray}
Here $\hat{A}^\dag$ and $\hat{A}$ stands for the creation and annihilation operators of the cavity field in the rotating frame at the pump frequency $\omega_p$, and $\eta_p$ is the cavity pumping strength. The motion of atoms with mass $m$ inside the traps with identical frequencies $\nu$ is described by the displacement operator $\hat{x}_i$ of the $i$-th atom from its trap center $x_i^{(0)}$. The single atom off-resonant coupling strength at the anti-node is $U_0=g^2/\Delta_a$.
The first term in the Hamiltonian contains the optomechanical interaction between the cavity field and the atomic motion: $U_0 \sum_{i=1}^{N}\cos^2(k_c x^{(0)}_i+k_c\hat{x}_i)$ is the shift of the cavity frequency caused by the presence of the atoms and, conversely, the mechanical potential exerted on the atoms by a single cavity photon. Further on only the Lamb-Dicke regime will be considered, when the atoms are localized on a length scale $\Delta x=\sqrt{\hbar/(2\nu m)}$ much smaller than the cavity wavelength $\lambda=2\pi/k_c$ ($\eta=k_c \Delta x$ is much smaller than one). Thus only the contributions up to the second order in the Lamb-Dicke parameter will be considered and the approximation ${\cos^2(k_c x^{(0)}_i+k_c\hat{x}_i)=\cos^2(k_c x^{(0)}_i)-\sin(2k_c x^{(0)}_i) k_c \hat{x}_i-\cos(2k_c x^{(0)}_i) (k_c \hat{x}_i)^2}$ will be used.

The incoherent dynamic due to the cavity decay and the spontaneous emission (up to the second order in $1/\Delta_a$) is captured by the following Heisenberg-Langevin equations:
\begin{eqnarray}
\label{H-L_equations}
\label{H-L_equations_sp_em}
\dot{\hat{A}}&=&\frac{i}{\hbar}\left[H,\hat{A}\right]
-(\kappa+\sum_{i=1}^ND_{ai}/2)\hat{A}
+\sqrt{2\kappa}\hat{S}_{a}
+i\sum_{i=1}^N\sqrt{D_{ai}}\hat{f}_{ai},
\nonumber
\\
\dot{\hat{p}}_i&=&\frac{i}{\hbar}\left[H,\hat{p}_i\right]
-2\Delta p\sqrt{D_{bi}}\hat{f}_{bi},\,\,\,\,\,\,
\dot{\hat{x}}_i=\frac{i}{\hbar}\left[H,\hat{x}_i\right].
\end{eqnarray}
where $\Delta p=\sqrt{\hbar\nu m/2}$. The above equations  are derived in the appendix by taking in to account the coupling of the atom-cavity system to the external electromagnetic environment and using the markovian approximation to eliminate the external field modes from the equation \cite{Bienert2012} prio to the elimination of the atomic internal degree of freedom \cite{Vitali2008}.

The noise operator $\hat{S}_a$ of the vacuum  field entering the cavity through the mirror has the zero mean value and its correlation functions are:
\begin{eqnarray}
\langle\hat{S}_{a}(t)\hat{S}^\dag_{a}(t')\rangle=\delta(t-t'),
\\
\nonumber
\langle\hat{S}^\dag_{a}(t)\hat{S}_{a}(t')\rangle=\langle\hat{S}_{a}(t)\hat{S}_{a}(t')\rangle=\langle\hat{S}^\dag_{a}(t)\hat{S}^\dag_{a}(t')\rangle=0.
\end{eqnarray}

The scattering of the cavity photons by the atoms in to the outer modes causes the Langevin forces $\hat{f}_{ai}$ and $ \hat{f}_{bi}$ correspond to the loss of the cavity photons with rate $D_{ai}$, and the diffusion of an atomic motion with rates $D_{bi}$ respectively. They have the following non-zero correlation functions:
\begin{eqnarray}
\label{eq_cor_fun_sp_em}
D_{ai}=\gamma \frac{g^2}{\Delta_a ^2}\cos ^2(k_cx^{(0)}_i),\,\,\,\,\,
D_{bi}=\gamma \frac{g^2}{\Delta_a ^2}\eta ^2 \alpha ^2K_i ,
\nonumber
\\
\langle \hat{f}_{ai}(t) \hat{f}^\dag_{ai}(t') \rangle = \delta(t-t'),
\nonumber
\\
\langle \hat{f}_{bi}(t)\hat{f}_{bi}(t') \rangle = \delta(t-t'),\,\,\,\,\,\,
\hat{f}^\dag_{bi}(t)= \hat{f}_{bi}(t),
\nonumber
\\
\langle \sqrt{K_i}\hat{f}_{bi}(t)\hat{f}_{ai}^\dag(t') \rangle =
\langle \hat{f}_{ai}(t) \sqrt{K_i}\hat{f}_{bi}(t')\rangle=\sin(k_cx^{(0)}_i)\delta(t-t').
\end{eqnarray}
Here $\alpha^2$ represents the mean cavity photon number in the zero order with respect to the
Lamb-Dicke parameter $\eta$. An order of unity coefficient $K_i=\sin^2(k_cx^{(0)}_i)+C_{xi}\cos^2(k_cx^{(0)}_i)$ depends on the atomic position along the cavity axes and on $C_{xi}= \int_{-1}^1 d\cos(\theta) \cos^2(\theta)\mathcal{N}_i(\cos(\theta))$, which gives the angular dispersion of the atom momentum and accounts for the dipole emission pattern $\mathcal{N}_i(\cos(\theta))$ \cite{SteckDataWeb}.  Equations (\ref{H-L_equations_sp_em}) and correlation functions (\ref{eq_cor_fun_sp_em}) are derived under the assumption that the inter-atomic distance $d$ is much larger that the cavity wavelength $k_cd\gg1$ which allows one to consider atoms as independent scatterers. 

In the case of a single atom, equations (\ref{H-L_equations_sp_em}) correspond to the result reported in \cite{Vitali2008} where the rates to raise and lower the vibration quanta also compensate each other up to the second order in $1/\Delta_a$ and only the diffusion effect remains. The difference with the result presented here accounts for the different pumping geometry - the atom is pumped from the side or the cavity is pumped through the mirror.

Next assumption on the way to solve equations (\ref{H-L_equations_sp_em}) is a large intracavity photon number with only small fluctuations around its steady state mean value: ${\langle\hat{A}^\dag\hat{A}\rangle\gg\langle\hat{a}^\dag\hat{a}\rangle}$, with ${\hat{a}=\hat{A}-\langle\hat{A}\rangle}$. The steady state mean values for the cavity field $\langle\hat{A}\rangle$, atom displacement $\langle\hat{x}_{i}\rangle$ and momentum $\langle\hat{p}_{i}\rangle$ are the solutions of the nonlinear algebraic equations constructed by taking the mean values on the left- and right-hand sides in equations (\ref{H-L_equations_sp_em}) and putting the derivatives to zero (assuming that the fluctuations are small) :
\begin{eqnarray}
\label{equations_mean_steady_state}
\langle\hat{A}\rangle=\frac{\eta_p}{(\kappa_\mathrm{eff}-i\Delta_c'-iU_0 \sum_{i=1}^{N}(s_ik_c\langle\hat{x}_{i}\rangle+
c_ik^2_c\langle\hat{x}_{i}^2\rangle)},
\nonumber
\\
k_c\langle\hat{x}_{i}\rangle=
\frac{2U_0 \eta |\langle\hat{A}\rangle|^2s_i}
{\nu-4U_0 \eta |\langle\hat{A}\rangle |^2c_i},\,\,\,\,\,
\langle\hat{p}_{i}\rangle=0.
\end{eqnarray}
Here $\Delta'_c=\Delta_c-U_0\sum_{i=1}^N\cos^2(k_cx_i^{0})$, $\kappa_\mathrm{eff}=\kappa+\sum_{i=1}^ND_{ai}/2$, and $c_i=\cos(2k_c x^{(0)}_i)$ and $s_i=\sin(2k_c x^{(0)}_i)$. Without any loss of generality we assume $\langle\hat{A}\rangle$ to be real, which can be adjusted by choosing the phase of $\eta_p$.  In the Lamb-Dicke regime the cavity mean field can be seen as a power series in the Lamb-Dicke parameter  $\langle\hat{A}\rangle=\alpha+O(\eta)$ with the zero order term
\begin{eqnarray}
\label{mean_cavity_field}
\alpha=\frac{\eta_p}{\kappa_\mathrm{eff}-i\Delta_c'}.
\end{eqnarray}

The evolution of small fluctuations around the steady state mean values is well described by the linear system of equations.Substituting $\hat{A}=\langle\hat{A}\rangle+\hat{a}$, $\hat{x}_i=\langle\hat{x}_{i}\rangle+\tilde{\hat{x}}_i $ and $\hat{p}_i=\langle\hat{p}_{i}\rangle+\tilde{\hat{p}}_i $ in to equations (\ref{H-L_equations_sp_em}) and neglecting the nonlinear terms together with other terms of the same order of magnitude brings us to the following equations:
\begin{eqnarray}
\label{H-L_equations_linearised}
\dot{\hat{a}}=\left(-\kappa_\mathrm{eff}+i\Delta_c'\right)\hat{a}+i\frac{U_0 \eta \alpha}{\Delta x}\sum_{i=1}^{N}s_i\tilde{\hat{x}}_i
+\sqrt{2\kappa}\hat{S}_{a}
+i\sum_{i=1}^N\sqrt{D_{ai}}\hat{f}_{ai},
\nonumber
\\
\dot{\tilde{\hat{p}}}_i=-m\nu_i^2\tilde{\hat{x}}_i+2\Delta p \,U_0 \eta \alpha\, s_i(\hat{a}^\dag+\hat{a})
-2\Delta p\sqrt{D_{bi}}\hat{f}_{bi},\,\,\,\,\,
\dot{\tilde{\hat{x}}}_i=\frac{\tilde{\hat{p}}_i}{m},
\end{eqnarray}
where $\nu^2_i=\nu(\nu-4 U_0 (\eta \alpha)^2c_i)$ is a modified trap frequency. To be consistent with the Lamb-Dicke approximation and to ensure that $|k_c\langle\hat{x}_{i}\rangle|\ll1$ in equation (\ref{equations_mean_steady_state}), the following inequality should be fulfilled:
\begin{equation}
\label{ineq_keep_LDL}
6 U_0 (\eta \alpha)^2\ll\nu.
\end{equation} 

Finally, linear equations (\ref{H-L_equations_linearised}) allow to reconstruct the effective Hamiltonian governing the dynamics of the cavity field fluctuations and atomic motion in the traps:
 \begin{eqnarray}
\label{Hammiltonian_effective}
H_\mathrm{eff}=-\hbar \Delta'_c \hat{a}^\dag \hat{a}
+\sum_{i=1}^{N}\left(\frac{m \nu_i^2 \hat{\tilde{x}}^2_i}{2}+\frac{\tilde{\hat{p}}^2_i}{2m}\right)
-\hbar \frac{U_0\eta \alpha}{\Delta x} (\hat{a}^\dag+\hat{a}) \sum_{i=1}^{N}s_i\hat{\tilde{x}}_i.
\end{eqnarray}
This expression shows the two main effects captured by our model: the optomechanical coupling responsible for the cooling mechanism (last term) and the modification of the trap frequencies as a mean field effect of the cavity potential. The trap inhomogeneity is an essential ingredient for cooling atoms to the ground state of motion. As only one collective mode of motion $\hat{X}\sim \sum_{i=1}^{N}s_i\tilde{\hat{x}}_i$ couples to the cavity, the cooling mechanism takes place exclusively by removing the excitations from this mode \cite{Schleier-Smith2011}. If the frequencies of all the traps are identical, for example when the inter-atomic distance in the array is a multiple of the cavity wavelength, this collective mode is also an eigenmode of a free atomic system.  Thus it will be decoupled from the remaining $N-1$ longitudinal modes of collective motion \cite{Genes2008}, and these modes will stay excited. As the steady state energy of each atom is determined by the weights of all the collective modes, individual atoms will be only partially cooled. Alternatively, if the trap frequencies are different than the collective mode $\hat{X}$ is no longer an eigenmode of a free atomic subsystem and it will be coupled to other $N-1$ longitudinal modes of collective atomic motion. This will allow a sympathetic cooling of all the collective modes, and all the atoms. The same principle is the basis for the sideband cooling of a trapped ion in three dimensions with a single laser beam \cite{Eschner2003}. In that case the requirements for cooling in all three dimensions are: different oscillation frequencies along each axes and a non-zero projection of the light wave vector on all the axes. The case of a one-dimensional cooling of many particles appears analogous to the cooling of a single particle in multiple directions. Similarly, in the case of an atomic array the conditions are: different trap frequencies and non-zero coupling of light to each atom.

\section{Analytical results: cooling rates}
\label{sec_anal_res}

This section is devoted to the analyses of the atom-cavity evolution neglecting the effect of the spontaneous emission ($\hat{f}_{ai}=\hat{f}_{ai}=D_{ai}=D_{bi}=0$) and taking in to account only the cavity decay. Importantly, this simplification will not significantly effect the cooling rates in the far off-resonance regime, provided that the cavity decay is much faster than the spontaneous emission rate:
\begin{equation}
\label{ineq_cavity_decay_vs_sp_em}
\kappa\gg \frac{\gamma g^2}{2\Delta_a^2}\sum_{i=1}^N\cos^2(k_cx^{(0)}_i),
\end{equation}
and $\kappa_{\mathrm{eff}}\approx \kappa$. In this regime the spontaneous emission will mainly cause a diffusion, the process in which the rate of adding and subtracting of a motion quantum are identical, and the contributions of both in to the final cooling rate cancel each other. Contrary the steady state phonon number for the atoms will increase due to the diffusion process and section \ref{sec_spont_em} will be devoted to this issue.

Direct cooling of collective mode $X$ and its exchange with the remaining collective modes may appear on different time scales and the slowest of them will correspond to the cooling time scale for individual atoms. This section presents the analytical limits for the cooling rates of different modes of motion and the scaling of the cooling dynamics with atom number $N$.

The cavity potential provides the trap inhomogeneity with a narrow distribution of the trap frequencies around $\nu$ in the Lamb-Dicke regime: ${\nu^2_i=\nu\left[\nu-4 U_0 (\eta \alpha)^2\cos(2k_cx^{(0)}_i)\right]}$. Distributing the atoms such that ${2k_cx^{(0)}_i=i\left(\frac{\pi}{N+1}+2n\pi\right)}$, $i=1,...N$, where $n$ is any integer will correspond to the following ratio between the inter-atomic distance $d$ and the cavity wavelength $\lambda$:
\begin{eqnarray}
\label{equation_periodicity}
\frac{d}{\lambda}=\frac{n}{2}+\frac{1}{4(N+1)}.
\end{eqnarray}
This configuration will simplify the calculation and will allow to find the analytical solutions for the cooling rates. More over, the results will capture the general properties of the cooling, regardless the atomic configuration.

It is convenient to introduce the collective modes in the following way: first mode $X$ coupled to the cavity and remaining modes $X_i$, $i=1,...N-1$, uncoupled from each other and coupled only to the first one. For the selected ratio $\frac{d}{\lambda}$ it can be done using the following transformation from the basis of individual atomic displacements:
 \begin{eqnarray}
\label{equations_collective_mode transformation}
X=\sqrt{\frac{2}{N+1}}\sum_{i=1}^N\sin(\frac{\pi \cdot i}{N+1})\tilde{\hat{x}}_i,
\\
\nonumber
X_i=\sqrt{\frac{2}{N}}\sum_{k=1}^{N-1}\sin(\frac{\pi \cdot i\cdot k}{N})\sqrt{\frac{2}{N+1}}\sum_{j=1}^N\sin(\frac{\pi \cdot j \cdot (k+1)}{N+1})\tilde{\hat{x}}_j.
\end{eqnarray}
Identical transformations relate the momenta of the collective modes $P$ and $P_i$ with the momenta of the individual atoms $\tilde{\hat{p}}_i$. Substituting this transformation into the effective Hamiltonian (\ref{Hammiltonian_effective}) and introducing the creation and anihilation operators $X=\sqrt{\frac{\hbar}{2m\nu}}(\hat{B}^\dag+\hat{B})$, $P=i\sqrt{\frac{\hbar m\nu}{2}}(\hat{B}^\dag-\hat{B})$, and $X_j=\sqrt{\frac{\hbar}{2m\omega_j}}(\hat{B}_j^\dag+\hat{B}_j)$, $P_j=i\sqrt{\frac{\hbar m\omega_j}{2}}(\hat{B}_j^\dag-\hat{B}_j)$ we get the Hamiltonian in the desired form:
\begin{eqnarray}
H_\mathrm{eff}=&-&\hbar\Delta'_c\hat{a}^\dag \hat{a}
       +\hbar\nu \hat{B}^\dag \hat{B}
       +\sum_{j=1}^{N-1}\hbar\omega_j \,\hat{B}_j^\dag  \hat{B}_j
\nonumber
\\
&-&\hbar\frac{\epsilon}{2}\left(\hat{a}+\hat{a}^\dag \right)(\hat{B}+\hat{B}^\dag)
-\hbar\sum_{j=1}^{N-1}\frac{\beta_j}{2}(\hat{B}+\hat{B}^\dag)(\hat{B}_j+\hat{B}_j^\dag).
\end{eqnarray}
The coupling strengths $\epsilon$, between the cavity and collective mode $X$, and $\beta_i$,  between collective modes $X$ and $X_i$, and the collective mode frequencies $\omega_i$ are:
\begin{eqnarray}
\label{eq_collective_couplings}
\epsilon=U_0\alpha\eta\sqrt{2(N+1)},
\nonumber
\\
\beta_j=2U_0(\alpha\eta)^2 \sqrt{\frac{2}{N}}\sqrt{\frac{\nu}{\omega_j}}\sin\left(\pi\frac{j}{N}\right),
\\
\nonumber
\omega^2_j=\nu\left[\nu-4U_0(\eta\alpha)^2\cos\left(\pi\frac{j}{N}\right)\right].
\end{eqnarray}
Thus the cavity coupling to the $X$ mode increases with $N$ while the coupling between $X_i$ and $X$ modes decreases with $N$. Such an opposite dependence will lead to the emergence of separate time scales when the atom number is sufficiently large. In this case the dynamic will consist of a fast excitation subtraction from the $X$ mode via the exchange with the cavity followed by the cavity decay, and a slow exchange between the modes $X_i$ and $X$.

To find separately the asymptotic expressions of the collective mode decay rates for the fast and slow processes for $N\gg1$, at first only the interaction between cavity and $X$ mode is considered ($\beta_i=0$). This single mode case was considered in the work \cite{Schleier-Smith2011} and it is analogous to the cavity cooling of a single trapped particle \cite{Vuletic2001,Zippilli2005} as well as of an eigenmode of a mechanical cantilever \cite{Marquardt2007}. It can be be described by the rate equations for a mean phonon number $N_X=\langle B^\dag B\rangle$ in the form: $\dot{N}_X=-\gamma_X\left(N_X-N_X(t\rightarrow\infty)\right)$ when the cavity mode is adiabatically eliminated \cite{Stenholm1986}.  In this work the rate equation is derived by evaluating $\langle\dot{\hat{B}^\dag\hat{B}}\,\,\rangle=\langle\dot{\hat{B}}^\dag\hat{B}\rangle+\langle\hat{B}^\dag\dot{\hat{B}}\rangle$. The cooling rate and the steady state phonon number are found to be:
\begin{eqnarray}
\label{eq_colective_cooling_rate}
\gamma_X=\frac{\epsilon^2}{2\kappa}\left[S_-(\nu) -S_+(\nu)\right]=\kappa\, c_d^2\,(\eta\alpha)^2(N+1)\left[S_-(\nu) -S_+(\nu)\right],
\\
\label{eq_colective_cooling_occupation_number}
N_X(t\rightarrow\infty)=\frac{S_+(\nu)}{S_-(\nu)-S_+(\nu)}.
\end{eqnarray}
 Here ${c_d=\frac{U_0}{\kappa}=\frac{g^2}{\Delta_a\kappa}}$ is an off-resonance single atom cooperativity which is the key parameter characterizing cavity-atom interaction and representing both: the cavity frequency shift due to the interaction with one atom and the atom resonance shift due to the interaction with the cavity in the units of the cavity lightweight. Spectral parameters ${S_\pm(\nu)=(1+(\Delta_c'\mp\nu)^2/\kappa^2)^{-1}}$ stand for the subtraction ($S_-$)/addition ($S_+$) of an energy quantum from/to the collective mode of motion and refer to the cooling and heating processes respectively. This description is applicable in the weak interaction regime when $\epsilon\ll\kappa$ which imposes an upper limit for the atom number in the array, $N\ll(c_d\eta\alpha)^{-2}$.

Efficient cooling of collective mode $X$ will occur at the cooling side-band $\Delta_c'=-\nu$ and in the resolved side-band regime $\kappa\ll\nu$. In this case $S_-(\nu)=1$ and $S_+(\nu)\approx\frac{\kappa^2}{4\nu^2}$ and the contribution of the heating processes is negligible. The cooling rate is then $\gamma_X\approx\epsilon^2/(2\kappa)=\kappa(c_d\eta\alpha)^2(N+1)$ while the mean phonon number $N_X(t\rightarrow\infty)\approx S_+(\nu)$ is close to zero. Assuming this regime, we now consider the evolution of the remaining modes. If the exchange between modes $X$ and $X_i$ occurs at the time scale much slower than $\gamma_X^{-1}$, mode $X$ will serve as a decay channel for the remaining modes.

We will look for the cooling rate for each $X_i$ mode independently, assuming that the effect of the presence of modes $X_j$ ($j\neq i$), can be neglected for sufficiently large $N$. We shall note that the condition similar to the one providing the resolved side-band regime is automatically fulfilled: the decay rate of mode $X$ is much smaller that the frequency of the $i$-th mode $\gamma_X\ll\omega_i$. Also the condition similar to the cooling side-band condition is fulfilled for each mode, $|\nu-\omega_i|\ll\gamma_X$ for sufficiently large atom number $(N+1)\gg2/c_d$. The independent cooling rate for the $X_i$ mode in this resolved side-band regime is well approximated by the following expression:
\begin{eqnarray}
\label{eq_colling_rates_analitical}
\gamma_{X_i} = \frac{\beta_i^2}{\gamma_X}=
\kappa \frac{8(\alpha\eta)^2}{(N+1)N}\sin^2\left(\frac{\pi\cdot i}{N}\right)\frac{\nu}{\omega_i}.
\end{eqnarray}
In the derivation of the rate equation for the mean phonon number in mode $X_i$, the counter-rotating terms $\hat{B}^\dag\hat{B}_i^\dag$ and $\hat{B}\hat{B}_i$ in the Hamiltonian were neglected (rotating wave approximation). Then, in the frame rotating with frequency $\nu$, the cavity mode $\hat{a}$ and collective mode $\hat{B}$ were subsequently eliminated to get the Heisenberg-Langevin equation for the $\hat{B}_i$ mode alone. The rate equation for the mean phonon number in each mode was again derived by calculating the mean value $\langle\dot{\hat{B}^\dag_i\hat{B}_i}\,\,\rangle=\langle\dot{\hat{B}}^\dag_i\hat{B}_i\rangle+\langle\hat{B}^\dag_i\dot{\hat{B}}_i\rangle$. The rotating wave approximation allows to reconstruct only the decay term but not the steady state mean phonon number in the rate equation because the heating side-band is neglected. Apart from this drawback it allows one to find the cooling rate with a good accuracy in the resolved sideband regime.

Expression (\ref{eq_colling_rates_analitical}) shows a non-linear decrease of the independent cooling rates with increasing atom number. The smallest rate which will determine the cooling rate of individual atoms is $\gamma_{X_1}\sim N^{-4}$ when the atom number is much bigger than one. Here we shall recall that while changing the atom number we keep the modified cavity detuning fixed to the cooling side-band $\Delta_c'=-\nu$ which means that the pump frequency is adjusted for each atom number such that $\Delta_c=-\nu+U_0\sum_{i=0}^Nc^2_i=-\nu+U_0(N-1)/2$. Also, the choice of the array periodicity (\ref{equation_periodicity}) made the ratio $d/\lambda$ dependent on $N$.

One should note that the cooling rates (\ref{eq_colling_rates_analitical}) in this collective cooling regime does not really depend on the interaction strength $U_0$, only weakly through $\omega_i$. It is at first surprising, but reasonable, since the cooling is a trade-off between two processes: the
exchange among different collective modes and the decay of the collective mode coupled
to the cavity. These two processes are initially governed by the interaction of the same
origin with strength $U_0$. When the collective mode decay rate increases the cooling slows down because less exchange events appear on the decay time scale, this is compensated by the
simultaneous growth of the exchange rate. Thus the single atom interaction strength
cancels out in the resulting cooling rate. This fact will crucially change the influence
of the spontaneous emission on the cooling process in comparison with the single atom
case which we will discuss in section \ref{sec_spont_em}.

\section{Numerical results: cooling rates and mean phonon numbers}
\label{sec_num_res}

The exact evolution of $N$ atoms coupled to the cavity mode described by Hamiltonian (\ref{Hammiltonian_effective}) cannot be found analytically and the cooling rates and mean phonon numbers are calculated numerically still neglecting the effect of the spontaneous emission 
 ($\hat{f}_{ai}=\hat{f}_{ai}=D_{ai}=D_{bi}=0$).
 
 Performing the transformations $\tilde{\hat{x}}_i=\sqrt{\frac{\hbar}{2m\nu_i}}(\hat{b}^\dag_i+\hat{b}_i)$, $\tilde{\hat{p}}_i=i\sqrt{\frac{\hbar m\nu_i}{2}}(\hat{b}^\dag_i-\hat{b}_i)$, where $\hat{b}^\dag$ and $\hat{b}$ are the creation and annihilation operators of a vibrational excitation for individual atoms, the system of equations (\ref{H-L_equations_linearised}) can be rewritten in the matrix form
\begin{equation}
\dot{Y}=MY+S.
\end{equation}
Here we introduced the vectors of the system fluctuations and the noise operators:
\begin{eqnarray}
Y=(\hat{a},\hat{b}^{}_1,\hat{b}^{}_2,...\hat{b}^{}_{N},\hat{a}^\dag,\hat{b}_1^\dag,\hat{b}_2^\dag,...\hat{b}_{N}^\dag)^T,
\\
\nonumber
S=(\sqrt{2\kappa}\hat{S}_{a},0,0,...0,\sqrt{2\kappa}\hat{S}^\dag_{a},0,0,...0)^T. 
\end{eqnarray} 
The dynamical matrix $M$ is non-Hermitian and its non-zero elements are $M_{aa}=M^*_{a^\dag a^\dag}=-\kappa+i\Delta_c'$, $M_{b_ib_i}=M^*_{b^\dag_ib^\dag_i}=-i\nu_i$ and $M_{a b_i}=M_{a b_i^\dag}=M^*_{a^\dag b_i}=M^*_{a^\dag b_i^\dag}=M_{b_i a}=M_{b_i a^\dag}=M^*_{b^\dag_i a}=M^*_{b^\dag_i a^\dag}=iU_0\eta\alpha s_i$. The transformation diagonalizing this matrix will result in the new operators combining light and atomic variables. The decay rates of the population of these polaritonic modes, $\Gamma_i$, are given by the real part of the eigenvalues $\mu_i$ of matrix $M$. Since a steady state energy of individual atoms are determined by the weighted energies of all the polaritonic modes, the smallest of $\Gamma_i$ will set the decay rate for individual atoms. 
 \begin{figure}
\begin{center}
\includegraphics[width=0.95\columnwidth]{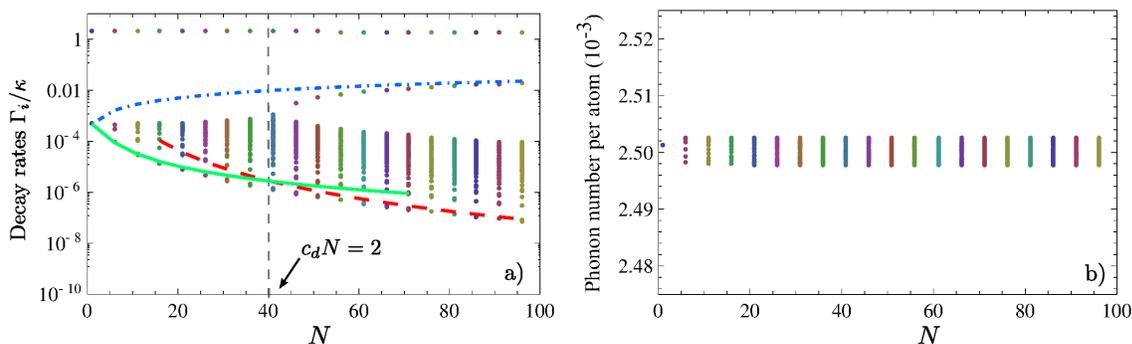}
\end{center}
\caption{Scaling with the atom number. For each $N$ there are $N+1$ points representing polaritonic decay rates $\Gamma_i=-2Re[\mu_i]$ (a) and $N$ points representing the phonon numbers per atom (b). Dashed and solid lines correspond to the analytical results: collective cooling rates $\gamma_X$ (\ref{eq_colective_cooling_rate}) (blue, dashed-dotted line) and $\gamma_{X_1}$ (\ref{eq_colling_rates_analitical}) (red, dashed line) and independent decay rate $\gamma_{x1}$ (\ref{eq_independent_cooling_rate}) (green, solid line). Vertical dashed line marks the transition between independent and collective cooling regimes. In both regimes the steady state mean phonon number per atom is close to the limit of a single atom cavity cooling $\kappa^2/(4\nu^2)=0.0025$. The cavity detuning is adjusted to the cooling sideband $\Delta_c'=-\nu$ for each atom number and the $\nu=10\kappa$. Parameter values $\eta=0.02$,  $c_d=0.05$,  $\eta_p=150\kappa$ result into $c_d(\eta\alpha)^2=4.8\cdot 10^{-3}$.}
\label{fig_scaling_all_rates}
\end{figure}

The decay rates of the polaritonic modes $\Gamma_i=-2Re[\mu_i]$, $i=1,...N+1$, are plotted in figure \ref{fig_scaling_all_rates}.a for different atom numbers in the array. The pump frequency was adjusted to keep the cooling sideband condition $\Delta_c'=-\nu$, and the atomic periodicity \emph{vs} the cavity wavelength $d/\lambda$ was also modified according to (\ref{equation_periodicity}). Other parameters are selected such that $\epsilon\ll\kappa$ and the cavity decay happens much faster than the phonon decay. In this case we clearly see the dominating decay rate $\Gamma_1\approx2\kappa$ corresponding to the polaritonic modes mainly consisting of the cavity mode. Consequently the remaining polaritonic modes will mostly consist of atomic modes. For sufficiently large $N$ the decay rates are well approximated by the analytical expressions for the collective mode decay rates $\gamma_X$ (blue, dashed-dotted line) and $\gamma_{X_i}$ (red, dashed line for  $\gamma_{X_1}$ of figure \ref{fig_scaling_all_rates}.a) thus these modes are close to the collective modes introduced in the previous section.

When the atom number is small, analytical results (\ref{eq_colective_cooling_rate},\ref{eq_colling_rates_analitical}) are no longer valid because the collective mode $X$ cannot be treated independently from the remaining modes $X_i$. It turns out that for $N\ll 2/c_d$, the polaritonic decay rates $\Gamma_i$ for $i=2,...N+1$ are well approximated by the independent decay rates of each atom (green, solid line for $\gamma_{x1}$), found by putting $s_j=0$ for $j\neq i$: 
\begin{eqnarray}
\label{eq_independent_cooling_rate}
\gamma_{x_i}=\frac{\epsilon_i^2}{2\kappa}\left[S_-(\nu_i) -S_+(\nu_i)\right]
=2\kappa\,c_d^2(\alpha\eta_i)^2\left[S_-(\nu_i) -S_+(\nu_i)\right].
\end{eqnarray}
Here we introduce an effective coupling strength $\epsilon_i=2U_0\eta_i\alpha s_i$ between the cavity mode and the motion of the $i$-th atom and a Lamb-Dicke parameter for each atom $\eta_i=k_c\sqrt{\hbar/(2m\nu_i)}$. From this we conclude that atoms do not feel the presence of each other and they are cooled down independently. This is due to the fact that the difference between the trap frequencies is larger than the mechanical damping rate of each atom $\gamma_{x_i}$ and there is no interference effect between the cooling of different atoms. On the contrary, for a large atom number the frequencies $\nu_i\approx\nu-2 U_0 (\eta \alpha)^2\cos(i\pi/(N+1))$ are close to each other and when the difference becomes smaller than $\gamma_{x_i}$ the light mediated interaction between the traps slows down the cooling. A similar interference effect was previously found for two mechanical modes of a micromirror in an optical cavity \cite{Genes2008}. In figure \ref{fig_scaling_all_rates}.a the transition point between the two regimes in the atomic array when one collective decay rate splits from the others is clearly seen. Its position depends on the array geometry and is captured by $c_dN=\mathrm{const}$ where $\mathrm{const}=2$ in the present configuration.  

The steady state mean occupation number of each atom, presented on \ref{fig_scaling_all_rates}.b, practically does not depend on the total number of atoms if the spontaneous emission is neglected. It is approximately the same for all atoms and it is close to the lowest value achievable for a single atom resolved side-band cooling $\kappa^2/(4\nu^2)$ (0.0025 for the selected parameters) when the diffusion due to spontaneous emission is negligible \cite{Vuletic2001}. This is due to the fact that the shifts of the trap frequencies are much smaller than the cavity bandwidth and the cooling sideband conditions are still fulfilled for all the atoms.

Comparison of the numerical and analytical results allows us to associate the collective modes of the atomic motion presented in the previous section with the normal polaritonic modes of the full system. It also revealed the transition between two different regimes when atoms are cooled independently or collectively.  Comparing the smallest collective decay rate $\gamma_{X_1}$ with the smallest independent decay rate $\gamma_{x_1}$ for $N\gg1$ we see the suppression by a factor $\gamma_{x_1}/\gamma_{X_1}=(c_dN/2)^2$. Thus, while the collective effects are favourable for the cooling of one mode shortening its cooling time linearly with $N$ \cite{Schleier-Smith2011},  they destructively suppress the cooling of individual atoms and prolong their cooling time quadratically with N.  

 \section{Optimal array periodicity \emph{vs} the cavity wavelength}
 \label{sec_optimization}
 
So far we analysed the configuration when the ratio between the lattice constant and the cavity wavelength was set by expression (\ref{equation_periodicity}), which corresponds to the spread of the trap frequencies over the whole available interval $\cos(i\pi/(N+1))\in(-1,1)$. This provides the largest frequency difference between the traps and supposedly fastest exchange between the collective modes.  However, in this case, atoms on the edge of the chain are weakly coupled to the cavity due to the factor $\sin(i\pi/(N+1))$, which slows down the cooling. This section shows the existence of the optimal configuration of atoms in the cavity which maximizes the cooling rate due to the trade-off between the frequency separations and the coupling to the cavity.
\begin{figure}
\begin{center}
\includegraphics[width=0.9\columnwidth]{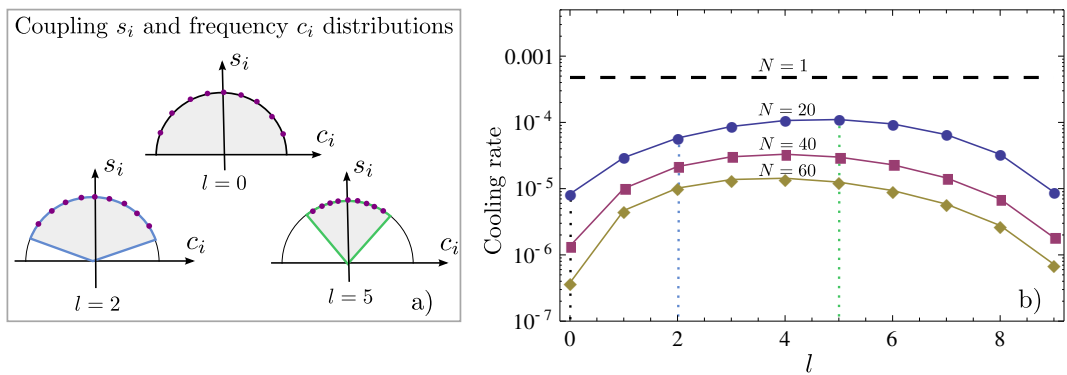}
\end{center}
\caption{Optimization of the cooling procedure. (a): Coupling strength $s_i=\sin(2k_cx_i^{(0)})$ and frequency $c_i=\cos(2k_cx_i^{(0)})$ distributions for 9 atoms and the optimization parameter $l=0,2,5$ with steps number $L=10$ (marked by the vertical dotted lines of figure \ref{fig_optimization}.b). (b): Minimal decay rate $\mathrm{Min}\{\Gamma_i\}$ \emph{vs} optimization parameter $l$. Blue circles, magenta squares and  yellow diamonds correspond to $N=20,40,60$ respectively. Reference dashed curve is a single atom cooling rate for $s_1=1$.}
\label{fig_optimization}
\end{figure}

Considering the frequency spread to be symmetric around $\nu$, the periodicity ratio $d/\lambda$ and the array location along the cavity axes will be varied to decrease the interval along which the trap frequencies are spread. This will automatically increase the minimal coupling to the cavity. Such a change can be parametrized as follows:
\begin{eqnarray}
2k_cx^{(0)}_i=\frac{l}{L}\cdot\frac{\pi}{2}+ i\left(\frac{L-l}{L}\cdot\frac{\pi}{N+1}+2n\pi\right),\,\,\,\,\, l=0,...L-1;
\\
\nonumber
\frac{d}{\lambda}=\frac{n}{2}+\frac{1}{4(N+1)}\cdot\frac{L-l}{L}.
\end{eqnarray} 
Here $L$ is the number of steps in the search for the optimal configuration. By changing the value of the optimization parameter $l$ from $0$ to $L-1$ we go from the largest to the smallest frequency spread. As an example, figure \ref{fig_optimization}.a shows the distribution of the frequencies  $c_i=\cos(2k_c x_i^{(0)})$ and couplings $s_i=\sin(2k_c x_i^{(0)})$  for nine atoms and the optimization parameter $l=0,2,5$ with $L=10$.

Figure \ref{fig_optimization}.b shows the change of the minimal cooling rate $\mathrm{Min\{\Gamma_i\}}$ with $l$ for three different atom numbers $N=20,40,60$. We find one order of magnitude improvement of the cooling rate as a result of the suggested optimization scheme. It is important to mention the role of the array location along the cavity axes. Displacement of the array away from the optimal location will be equivalent to the rotation of the selected segment shown on figure \ref{fig_optimization}.a around the origin. This would lead to the reduction of the cooling rate fro some atoms due to the decrease of the coupling to the cavity. Additionally if some $c_i$ become identical, some collective mode of motion will decouple and consequently the steady state phonon number per atom will increase.

It is experimentally convenient that the trap frequency inhomogeneity is provided by the cavity potential itself because no extra arrangements are needed to lift the trap degeneracy. Additionally, the key role of the array \emph{vs} cavity field periodicity may be used to speed up the cooling by a factor ${\sim N^2}$ by only displacing and stretching the array along the cavity. Alternatively additional external potentials can be considered to introduce an arbitrary trap inhomogeneity, however this is beyond the scope of this paper.

\section{Effect of spontaneous emission}
\label{sec_spont_em}

Up to now only the exchange between the atoms and the cavity mode was considered, and the spontaneous emission of the cavity photons by the atoms into the free space was neglected. Spontaneous emission on a single atom causes diffusion \cite{Vuletic2001,Bienert2012,Zippilli2005a} and, thus, heating. This leads to a higher steady state phonon number than predicted by the model neglecting the spontaneous emission. Now it will be take in to account by considering the additional Langevin sources and decay terms in equations (\ref{H-L_equations_linearised}) which were omitted in the previous sections. In general, the many atom case is different from the single atom configuration. Nevertheless, we can already guess that in the individual cooling regime, when the atom number is sufficiently small, the many- and single-atom cases will be similar and here it will be proven  analytically. More importantly, in this section I will also treat in detail the effect of the spontaneous emission in the collective cooling regime. We will see that the destructive suppression of the cooling rates discussed in the previous sections leads yet to another problem when accounting for the spontaneous emission: as the cooling slows down, the diffusion due to the spontaneous scattering into the free space accumulates during a longer time. This increases the steady state photon number in the traps setting an additional limitation for the proposed cooling scheme. This section presents both the numerical and analytical studies of the effect including the results derived for the first time for the considered configuration: many atoms in a pumped cavity. The results will be used to find the guidelines on how to set the parameters to avoid undesirable heating and to achieve the proposed cooling scheme experimentally feasible.

In the regime of the independent cooling, i.e. when the atom number is sufficiently small, we shall compare the exact solution with the analytical result for a single atom. The rate equation for the mean occupation number in $i$-th trap is derived by putting $\epsilon_ j$ ($j\neq i$) to zero and adiabatically eliminating the cavity mode assuming that $\kappa_\mathrm{eff}\gg\epsilon_i$. The cooling rate remains the same (\ref{eq_independent_cooling_rate}) and the steady state phonon number is found to be:
\begin{equation}
\label{eq_phonon_number_sp_em}
n_{i}(t\rightarrow\infty)=
\frac{S_+(\nu_i)}{S_-(\nu_i)-S_+(\nu_i)}\left(1
+\frac{1}{2c_r}\frac{K_i}{s^2_i}\frac{1}{S_+(\nu_i)}\right).
\end{equation}
In the expression for $S_\pm(\nu_i)$ the cavity decay rate $\kappa$ should be replaced by the modified rate $\kappa_\mathrm{eff}$, although under the condition (\ref{ineq_cavity_decay_vs_sp_em}) the dominating effects of the spontaneous emission will be captures if $\kappa_\mathrm{eff}\approx\kappa$, so will be assumed in the following. This expression clearly demonstrates the necessity of a large cooperativity $c_r$ to reach the ground state cooling. It is in agreement with the results reported in \cite{Vuletic2001,Zippilli2005a} with the only difference being a numerical factor of the order of unity accounting for different pumping configuration. This result also coincides (up to the second order in $1/\Delta_p$) with the the result reported in \cite{Bienert2012}, there the cavity pump configuration was also considered. It is interesting to note that the initial assumption on the intra-cavity mean photon number made in this work ($|\alpha|^2\gg\langle\hat{a}^\dag\hat{a}\rangle$) is essentially different to the one of \cite{Bienert2012} ($|\alpha|^2\ll1$). The exact agreement between the results underlines that the limit of a small intra-cavity photon number and the limit of a small fluctuation around a large inta-cavity photon number are two related approximations in the far off-resonance regime.
  
\begin{figure}
\begin{center}
\includegraphics[width=0.95\columnwidth]{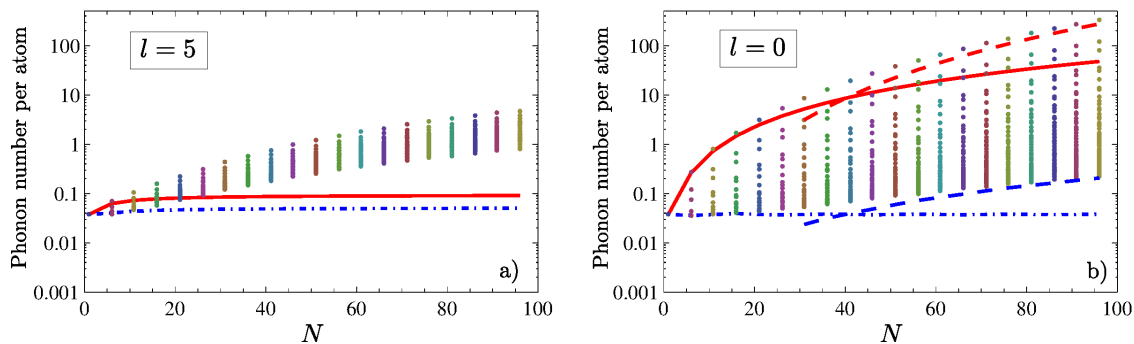}
\end{center}
\caption{Steady state occupation numbers per atom vs the atom number N. (a): optimal configuration with $l=5$, $L=10$, (b): configuration corresponding to figure \ref{fig_scaling_all_rates} with $l=0$. A resonance cooperativity $\frac{g^2}{\kappa\gamma}=10$ and other parameters are identical to those of figure \ref{fig_scaling_all_rates}. Analytical result for a single atom (\ref{eq_phonon_number_sp_em}) reproduces the numerical calculation for small $N$ ($N$-th trap (red, solid line) and  first (a) or middle (b) trap of the array (blue, dot-dashed line)). For the large atom number the numerical results are reproduced by expressions (\ref{eq_phonon_number_col_sp_em}) found in the collective cooling regime (red and blue dashed lines in (b)). The geometric coefficient $C_{xi}$ is set to 2/5 (the case of a classical dipole parallel to $x$ axes).}
\label{fig_occupation number_decay_rates_sp_em}
\end{figure}

For the case of $N$ atoms inside a cavity the problem is now solved numerically and the results are compared  with the a  single atom case in figure \ref{fig_occupation number_decay_rates_sp_em}. The scaling of the steady state phonon number with $N$ is presented for two different atomic configurations: the optimized configuration (figure \ref{fig_occupation number_decay_rates_sp_em}.a) and the one considered in figure \ref{fig_scaling_all_rates} (figure \ref{fig_occupation number_decay_rates_sp_em}.b). In agreement with the cooling rate scaling presented on figure \ref{fig_scaling_all_rates} the steady state phonon number scaling confirms that up to a certain atom number atoms cool down independently according to (\ref{eq_phonon_number_sp_em}). Above this atom number the cooling slows down which causes the increase of the phonon number as more spontaneous emission events occur during a longer cooling time. Thus  the transition from the individual to the collective cooling accompanied by the suppression of the cooling rate and the increase of the mean phonon number quadratically with the atom number is present in both configuration. For the selected parameters $c_d=0.05$ and $c_r=10$ up to $20$ atoms can be cooled close to the ground state with the phonon number less than $0.1$.

In the resolved side band limit $\kappa\ll\nu$ expression (\ref{eq_phonon_number_sp_em}) simplifies towards ${n_{i}\approx
\frac{\kappa^2}{4\nu_i^2}
+\frac{1}{2c_r}\frac{K_i}{s_i^2}\left(1+\frac{\kappa^2}{4\nu_i^2}\right)}$ and it is possible to estimate the steady state phonon number in the regime of collective cooling by taking into account the ratio between the individual (\ref{eq_independent_cooling_rate}) and collective (\ref{eq_colling_rates_analitical}) cooling rates $\gamma_{x_1}/\gamma_{X_1}=(c_dN/2)^2$:
\begin{equation}
\label{eq_phonon_number_col_sp_em}
n_{i}(N\gg 2/c_d)=
\frac{\kappa^2}{4\nu_i^2}
+\frac{(c_dN/2)^2}{2c_r}\frac{K_i}{s_i^2}\left(1+\frac{\kappa^2}{4\nu_i^2}\right).
\end{equation}
As can be seen in figure \ref{fig_occupation number_decay_rates_sp_em}.b, this expression reproduces the exact result for the atom number $N\gg 2/c_d$ .
To suppress the spontaneous emission effect (the second term) the single atom cooperativity should obey the inequality:
\begin{equation}
\label{ineq_cooperativity_sp_em}
c_r\gg c_d^2 N^2/(8 s_i^2).
\end{equation}
This is fundamentally different from the condition in the case of a single atom $c_r \gg1$
where $c_d$ does not enter and consequently the detuning does not play a role. It is because the cooling rate (\ref{eq_colling_rates_analitical}) no longer depends on $c_d$ and thus on the detuning, while the spontaneous emission rate does. As we see from (\ref{ineq_cooperativity_sp_em}), in the case of collective cooling the cooperativity $c_r$ is required to be larger than in a single atom, i.e. the positive effect of the cavity is corrupted by the destructive interference in the cooling dynamic.
But at the same time the detuning is becoming a knob to reduce the diffusion caused by
the spontaneous emission.

Inequality (\ref{ineq_cooperativity_sp_em}) is equivalent to ${\kappa\gg\gamma \frac{g^2}{\Delta_a^2}N^2/(8 s_i^2)}$.  The optimization decreases the phonon number for the hottest atom ($s_i\approx \pi/N$) and improves the scaling by a factor ${\sim N^2}$. In this case the condition sufficient to suppress the effect of spontaneous emission is found to be:
\begin{equation}
\label{ineq_supression_sp_em}
{\kappa\gg\gamma \frac{g^2}{\Delta_a^2}N^2}.
\end{equation}
This inequality should be compared to the condition (\ref{ineq_cavity_decay_vs_sp_em}) insuring that the spontaneous emission rate is much lower than the cavity decay rate, ${\kappa\gg\gamma \frac{g^2}{\Delta_a^2}N}$, assumed through the derivations. Condition (\ref{ineq_cavity_decay_vs_sp_em}) was also considered to be sufficient for neglect the spontaneous emission effect in the configuration different to the one presented here, i.e. homogeneour cold atomic cloud instead of the array \cite{Gangl1999,Horak2000}. As we can see, an additional factor of $N$ makes condition (\ref{ineq_supression_sp_em}) more strict than (\ref{ineq_cavity_decay_vs_sp_em}). This is a special feature of the collective cooling regime, when the distructive interference suppresses the cooling effect.

Lets now estimate experimental accessibility of the proposed cooling scheme for a chain of $^{87}$Rb atoms using the limitation (\ref{ineq_supression_sp_em}) as a guideline. Given the recoil frequency $\omega_R=2\pi\cdot3.9$ kHz and demanding a Lamb-Dicke parameter of $\eta=0.04$, the trap frequencies shall be set to $\nu=2\pi\cdot 2.4$ MHz. The resolved side-band condition requires the cavity bandwidth to be at maximum $\kappa=2\pi\cdot240$ kHz. From figure \ref{fig_optimization}.a the cooling rate for the array of 20 atoms is $10^{-4}\kappa$ which gives a cooling time of about $6.6$ ms. This is a realistic time comparable with the stability of an optical trap which will form the array, and it is close to the single atom cooling time experimentally achieved via Raman side-band cooling \cite{Reiserer2013}. This rate can be achieved with the single atom-cavity coupling strength $g=2\pi\cdot 3.8$ MHz leading to cooperativity $c_r=10$ and the detuning from the atomic resonance $\Delta_a=2\pi\cdot 1.2$ GHz. The diffusion due to the spontaneous emission will set the limit for the number of atoms which can be cooled to the ground state. The upper bound of this limit can be estimated form condition (\ref{ineq_supression_sp_em}), ${N^2\ll\kappa\left(\gamma\frac{g^2}{\Delta_a^2}\right)^{-1}}$, and it is about ten for the selected parameters. To push this limit without changing the cooling rate constant one could go further away from the atomic transition and simultaneously increase the coupling strength $g\sim \sqrt{\Delta_a}$. 

The cavity cooling protocol for an atomic array proposed in this work is shown to be limited by the presence of spontaneous emission. The Heisenberg-Langevin equations derived for the first time in the considered configuration were used to quantify this limitations and shown that the proposed scheme is experimentally feasible. Moreover the predicted cooling times for an array of tens of atoms at the reachable experimental is comparable with the best achived up to date for a single atom case  \cite{Reiserer2013}.

\section{Conclusion}
\label{sec_conclusion}

Cooling of the array of an atomic array via coupling to a single mode cavity is accessible when the inhomogeneity of the atomic trap frequencies is present. This work shows that the intra-cavity field with sufficiently large photon number is able to provide this inhomogeneity, simultaneously mediating the cooling of atoms to the ground state of the individual wells.

The cooling dynamics drastically changes with the size of the array from (i) the regime when atoms are cooled independently from each other to (ii) when the cooling happens via collective modes which increase the cooling time and the steady state mean phonon number by a factor $\sim(c_dN)^2$. The main reason for the suppression of the cooling at the large atom number is the destructive interference occurring because the separations between the trap frequencies become comparable with mechanical damping rate (an analog of the linewidth). It results into the destructive suppression of the cooling which is a signature of an enhancement of the cavity mediated atom-atom interaction. Consequently the detrimental spontaneous emission effect increases with the atom number and a larger single atom cooperativity $c_r\gg (c_dN)^2$ is necessary to suppress it.

Due to the periodic nature of the inhomogeneity induced by the cavity field the periodicity of the array \emph{vs} the cavity mode plays a crucial role in the cooling dynamics. It allows an optimization of the cooling by adjusting the lattice constant and the array position along the cavity axes offering one order of magnitude gain in the cooling speed. Cooling of a few tens of atoms to the ground state of motion within a few milliseconds is experimantaly feasible with the use of the suggested scheme. This demonstrates a controlability of the array motion with a single mode cavity and  sets the basis for the further exploration of the quantum optomechanical interface and, possibly, generation of novel non-classical states of collective atomic motion. Moreover, our cooling scheme can also be extended to the case of an array of micro- or manometer scale mechanical oscillators which makes it a useful tool for different systems.

\ack

I would like to thank Giovanna Morigi for the wise guidelines along the project and comments
on the manuscript, Marc Bienert for the discussions, critical reading and comments on the
manuscript, Cecilia Cormick, Endre Kajari, and Monika Schleier-Smith for the fruitful
discussions of the work in progress, Thomas Fogarty for reading the manuscript and for the
linguistic advising, and Pavel Bushev and Lars Madsen for the useful comments on the
manuscript. The research leading to these results has received funding from the  European Union Programme (FP7/2007-2013) under the FP7-ICT collaborative project AQUTE (grant number: 247687) and the individual Marie Curie  IEF project AAPLQIC (grant number: 330004). 

\begin{appendix}
\section{Heisenberg-Langevin equations: atom motion and cavity light }

The main ideas and the key steps of the derivation of equations (\ref{H-L_equations_sp_em}) and (\ref{eq_cor_fun_sp_em}) are presented in this appendix. The starting point is the full Hamiltonian of the system, which includes the cavity field, the atoms with their spin and mechanical degrees of freedom, and the reservoir containing the field modes outside of the cavity which interact directly with the atoms:
\begin{eqnarray}
\mathrm{H}_\mathrm{tot}=\mathrm{H}_\mathrm{sys}
+\sum_{\vec{k},\epsilon}\hbar\omega_{k}\hat{a}^\dag_{\vec{k},\epsilon}\hat{a}_{\vec{k},\epsilon}
- \sum_{
{ i=1} \atop {\vec{k},\epsilon}
}
    ^{N}
    \hbar g_{\vec{k},\epsilon}
                            \left(
                            \sigma^{(i)}_{eg}\hat{a}_{\vec{k},\epsilon} e^{i\vec{k}\vec{\hat{r}}_i}
                        +\sigma^{(i)}_{ge}\hat{a}^\dag_{\vec{k},\epsilon}e^{-i\vec{k}\vec{\hat{r}}_i}
                 \right).
\end{eqnarray}
The creation  $\hat{a}^\dag_{\vec{k},\epsilon}$ and annihilation $\hat{a}_{\vec{k},\epsilon}$ operators of the reservoir modes are labeled by the wave vector $\vec{k}$ and the polarization $\epsilon$ indexes and the summation goes over all the free space modes excluding those entering  through the cavity mirrors. The last term in the Hamiltonian represents the interaction between the atoms and the reservoir field modes in the rotating wave approximation with the interaction constant $g_{\vec{k},\epsilon}=\sqrt{\frac{\omega_k}{2 \pi \hbar V \epsilon_0}}\,\,\,\vec{\varepsilon}_\epsilon\cdot\vec{d}_{eg}$. Here $\vec{d}_{eg}$ is an atomic dipole moment, $V$ is the quantization volume and $\epsilon_0$ the vacuum permittivity. The spin of the $i$-th atom is represented by the operators $\sigma^{(i)}_{ge}=|g\rangle_i\langle e|$, $\sigma^{(i)}_{eg}=|e\rangle_i\langle g|$ and $\sigma^{(i)}_{z}=|g\rangle_i\langle g|-|e\rangle_i\langle e|$. The atom-cavity Hamiltonian $\mathrm{H_{sys}}$ contains the non-interacting parts $H_\mathrm{0}$, the interaction part $H_\mathrm{int}$ and the cavity pumping $H_\mathrm{p}$ :
\begin{eqnarray}
\label{eq_system_Hamiltonian}
\mathrm{H}_\mathrm{sys}=H_\mathrm{0}+H_\mathrm{int}+H_\mathrm{p},
\nonumber
\\
H_\mathrm{0}=-\hbar\frac{\omega_{eg}}{2} \sum_{i=1}^{N}\sigma^{(i)}_z-\hbar \omega_c \hat{A}^\dag \hat{A}
+\sum_{i=1}^N \left( \frac{m \nu^2}{2}\hat{x}^2_i+\frac{1}{2m}\hat{p}^2_i\right),
\nonumber
\\
H_\mathrm{int}=-\hbar g\sum_{i=1}^{N} \cos(k_c x^{(0)}_i+k_c\hat{x}_i)\left(\sigma^{(i)}_{eg} \hat{A}+\hat{A}^\dag \sigma^{(i)}_{ge}\right)
\nonumber,
\\
H_\mathrm{p}=i\hbar\left(\eta_p\hat{A}^\dag e^{-i\omega_p t}-\eta_p^*\hat{A}e^{i\omega_p t}\right).
\end{eqnarray}
The Heisenberg-Langevin equations for the atom-cavity system and reservoir are:
\begin{eqnarray}
\label{eq_HL_system+reservoir}
\dot{\hat{a}}_{\vec{k},\epsilon}=-i \omega_k \hat{a}_{\vec{k},\epsilon}
                                      - g_{\vec{k},\epsilon}\sigma^{(i)}_{ge} 
                                        e^{-i\vec{k}\cdot\vec{\hat{r}}_i},
\\
\nonumber
\dot{\hat{A}}=\frac{i}{\hbar}\left[\mathrm{H_{sys}},\hat{A}\right]
-\kappa \hat{A}
+\sqrt{2\kappa}\hat{A}_{in},
\\
\nonumber
\dot{\sigma}^{(i)}_{ge}=\frac{i}{\hbar}\left[\mathrm{H_{sys}},\sigma^{(i)}_{ge}\right]
+i \sigma^{(i)}_{z}\sum_{\vec{k},\epsilon}
                   g_{\vec{k},\epsilon}\hat{a}_{\vec{k},\epsilon}
                   e^{i\vec{k}\cdot\vec{\hat{r}}_i},
\\
\nonumber
\dot{\sigma}^{(i)}_z=\frac{i}{\hbar}\left[\mathrm{H_{sys}},\sigma^{(i)}_{z}\right]
+2i\sum_{\vec{k},\epsilon}
                   \left(
                   g^*_{\vec{k},\epsilon}\sigma^{(i)}_{ge}\hat{a}^\dag_{\vec{k},\epsilon}
                   e^{-i\vec{k}\cdot\vec{\hat{r}}_i}
                         -g_{\vec{k},\epsilon}\sigma^{(i)}_{eg}\hat{a}_{\vec{k},\epsilon}
                   e^{i\vec{k}\cdot\vec{\hat{r}}_i}
                   \right),
\\
\nonumber
\dot{\hat{p}}_i=\frac{i}{\hbar}\left[\mathrm{H_{sys}},\hat{p}_i\right]
+i\sum_{\vec{k},\epsilon}
                  \hbar k_x
                   \left(
                  g^*_{\vec{k},\epsilon}\sigma^{(i)}_{ge}
                                          \hat{a}^\dag_{\vec{k},\epsilon}
                                           e^{-i\vec{k}\cdot\vec{\hat{r}}_i}
                 -g_{\vec{k},\epsilon}\sigma^{(i)}_{eg}
                                           \hat{a}_{\vec{k},\epsilon}
                                            e^{i\vec{k}\cdot\vec{\hat{r}}_i}
                   \right),
\\
\nonumber
\dot{\hat{x}}_i=\hat{p}_i/m.
\end{eqnarray}
The first step on the way to the equation which contain only the atomic quantum motion and the cavity field is to eliminate the reservoir. It is done by formally solving the first equation of system (\ref{eq_HL_system+reservoir}):
\begin{eqnarray}
\hat{a}_{\vec{k},\epsilon}(t)=\hat{a}_{\vec{k},\epsilon}(0)e^{-i\omega_k t}
                       -ig^*_{\vec{k},\epsilon}
                       \int_0^t e^{-i\omega_k(t-\tau)}\sum_{i=1}^N \sigma^{(i)}_{ge}(\tau)
                                     e^{-i\vec{k}\cdot\vec{\hat{r}}_i}d\tau,
\end{eqnarray}
and plugging this solution into the remaining equation of system  (\ref{eq_HL_system+reservoir}). Assuming a markovian memoryless reservoir \cite{Cohen-Tannoudji1992,Gardinner2004} the system of equations can be developed to the following form:
\begin{eqnarray}
\label{eq_HL_spin+motion+losses}
\dot{\sigma}^{(i)}_{ge}=\frac{i}{\hbar}\left[\mathrm{H_{sys}},\sigma^{(i)}_{ge}\right]
-\frac{\gamma}{2}\sigma^{(i)}_{ge}
+i \sigma^{(i)}_{z}\hat{F}_i(t),
\\
\nonumber
\dot{\sigma}^{(i)}_z=\frac{i}{\hbar}\left[\mathrm{H_{sys}},\sigma^{(i)}_{z}\right] 
         -\gamma \left(\sigma^{(i)}_z+\mathrm{I}\right)
        +2i\left(
             \hat{F}_i^\dag(t) \sigma^{(i)}_{ge}
             -\sigma^{(i)}_{eg}\hat{F}_i(t)
              \right),
\\
\nonumber
\dot{\hat{p}}_i=\frac{i}{\hbar}\left[\mathrm{H_{sys}},\hat{p}_i\right]
+i\left(
             \hat{F}_{pi}^\dag(t) \sigma^{(i)}_{ge}
             -\sigma^{(i)}_{eg}\hat{F}_{pi}(t)
              \right),
\end{eqnarray}
with $\mathrm{I}$ is the identity operator. 
Here the Langevin sources contain the operators
${\hat{F}_i(t)=\sum_{\vec{k},\epsilon}g_{\vec{k},\epsilon} 
											e^{i\left(\vec{k}\cdot\vec{\hat{r}}_i(t)-\omega_k t\right)}                                           
                                            \hat{a}_{\vec{k},\epsilon}(0)}$ and  
${\hat{F}_{pi}(t)=\sum_{\vec{k},\epsilon}\hbar k_xg_{\vec{k},\epsilon} 
											e^{i\left(\vec{k}\cdot\vec{\hat{r}}_i(t)-\omega_k t\right)}                                           
                                            \hat{a}_{\vec{k},\epsilon}(0)}$
 accounting for the noise entering the atom-cavity system from the reservoir. Under the assumption that all the modes of the reservoir are in the vacuum state the only non zero correlation function of the operators $\hat{F}_i$ and $\hat{F}_i^\dag$ is ${\langle\hat{F}_i(t)\hat{F}_i^\dag(t')\rangle=g(t-t')e^{-i\omega_{eg}(t-t')}}$. The function ${g(\tau)=\sum_{\vec{k},\epsilon}|g_{\vec{k},\epsilon}|^2e^{-i(\omega_k-\omega_{eg})\tau}}$ is not exactly the delta-function although if the reservoir bandwidth is much larger than the inverse of the smallest time step considered in the problem then, it approaches a delta function ${\int_{-\infty}^{+\infty} g(\tau)d\tau=2\pi\sum_{\vec{k},\epsilon}|g_{\vec{k},\epsilon}|^2\delta(\omega_k-\omega_{eg})=\gamma}$ \cite{Cohen-Tannoudji1992}. Apart from the spontaneous decay rate $\gamma$ also a negligibly small energy shift additional to $\omega_{eg}$ appears due to the spontaneous emission which will further on be reabsorbed into the frequency.     
                                            
The second step is the adiabatic elimination of the atomic excited state in the limit of the large detuning $\Delta_a\gg\gamma,g\sqrt{N_{ph}},\kappa,\nu$. This is done by formaly solving the first two equations of system (\ref{eq_HL_spin+motion+losses}) and expanding the solution up to the second order in $1/\Delta_a$:
\begin{eqnarray}
\label{eq_optical_coherence_2order}
\sigma^{(i)}_{ge}=
&-&\frac{g f(\hat{x}_i)}{\Delta_a}\left[
\left(1+\frac{\Delta_c}{\Delta_a}+i\frac{\frac{\gamma}{2}-\kappa}{\Delta_a}\right)\hat{A}^\dag
+\frac{i}{\Delta_a}\left(\eta_p^*e^{i\omega_p t} + \sqrt{2\kappa}\hat{A}^\dag_{in}\right)
\right]
\\
\nonumber
&-&i\frac{g\sqrt{\omega_R\nu}}{\sqrt{2}\Delta^2_a}f'(\hat{x}_i)\frac{\hat{p}_i}{\Delta p}\hat{A}^\dag
+i\frac{\hat{F}^\dag(t)}{i\Delta_a+\gamma/2}
+O(\frac{1}{\Delta_a^3}).
\end{eqnarray}
The geometric functions depending on the positions of the atoms along the cavity are $f(\hat{x}_i)=\cos(k_cx_i^{(0)})-\sin(k_cx_i^{(0)})k_c\hat{x}_i-\frac{1}{2}\cos(k_cx_i^{(0)})(k_c\hat{x}_i)^2$ and   $f'(\hat{x}_i)=-\sin(k_cx_i^{(0)})-\cos(k_cx_i^{(0)})k_c\hat{x}_i$ up to the second order the Lamb-Dicke parameter.

The final point needed to arrive from equations (\ref{eq_HL_spin+motion+losses}) to equation (\ref{H-L_equations_sp_em}) and (\ref{eq_cor_fun_sp_em}) is the relations between the functions $\hat{F}_i(t)$ and  $\hat{F}_{p_i}(t)$ and the normalized Langevin sources  $ \hat{f}_{ai}(t)$ and $ \hat{f}_{bi}(t)$:
\begin{eqnarray}
\hat{f}_{ai}(t)=\frac{1}{\gamma}\hat{F}_i e^{i\omega_p t}
\\
\nonumber
\hat{f}_{bi}(t)=\frac{\cos(k_c x_i^{(0)})}{\sqrt{\gamma}}
                             \left(\hat{F}^\dag_i e^{-i\omega_p t}+\hat{F}_i e^{i\omega_p t}\right)
                       +\frac{i\sin(k_c x_i^{(0)})}{\sqrt{\gamma}\hbar k_c}
                             \left(\hat{F}^\dag_{pi} e^{-i\omega_p t}-\hat{F}_{pi} e^{i\omega_p t}\right)
\end{eqnarray}
Equations (\ref{H-L_equations_sp_em}) and (\ref{eq_cor_fun_sp_em}) are then derived from equations (\ref{eq_HL_spin+motion+losses}) using these relations, expression (\ref{eq_optical_coherence_2order}) and keeping only the terms up to the second order in $1/\Delta_a$.

\end{appendix}

\end{document}